\documentclass{article}
\usepackage[utf8]{inputenc} 
\usepackage{graphicx} 
\usepackage{slashed,epsfig,amsmath,amssymb,enumitem} \usepackage[papersize={8.5in,11in}]{geometry}
   \usepackage{bm}
\usepackage{graphicx}
\newcommand{\be}{\begin{equation}}
\newcommand{\ee}{\end{equation}}
\title{CMB Acoustic Power Spectra in STVG-MOG}
\author{J. W. Moffat\\
Perimeter Institute for Theoretical Physics, Waterloo, Ontario N2L 2Y5, Canada\\
and\\
Department of Physics and Astronomy, University of Waterloo, Waterloo,\\
Ontario N2L 3G1, Canada}
\begin{document}
\maketitle

\begin{abstract}
We present a cosmological realization of Scalar--Tensor--Vector Gravity (STVG--MOG) in which the pre-recombination scalar perturbation dynamics become degenerate with those of $\Lambda$CDM without invoking particle dark matter. In the early universe, nonrelativistic excitations of the massive STVG vector field $\phi_\mu$ behave as a collisionless, pressureless component with vanishing sound speed and background density $\rho_\phi \propto a^{-3}$. On the Fourier scales relevant for the acoustic peaks, the effective gravitational coupling satisfies $G_{\rm eff}(k,a)\simeq G_N$, so that the metric potentials governing baryon--photon oscillations evolve in the same way as in the standard cosmological model. The gravitational wells remain sufficiently deep at horizon entry to preserve the observed height of the third acoustic peak, the most sensitive indicator of a clustering pressureless component prior to recombination. Since Thomson scattering, recombination, baryon loading, and photon diffusion are unchanged, the temperature and polarization spectra can coincide with the standard $\Lambda$CDM predictions once the vector sector supplies the effective dust component. In this framework, the dynamical role usually attributed to cold dark matter is carried instead by a degree of freedom belonging to the gravitational sector itself. We explain why this vector-sector dust, although dynamically degenerate with cold dark matter in the early universe, is not equivalent to a particle dark matter fluid. The Boltzmann code CLASS is used to obtain a MOG fit to the acoustical power spectrum data.
\end{abstract}

\section{Introduction}

The angular power spectra of the cosmic microwave background (CMB) are among the most precise probes of the physics of the early universe. Measurements by the Planck satellite, together with high-resolution observations from the Atacama Cosmology Telescope (ACT) and the South Pole Telescope (SPT), have determined the temperature and polarization spectra with remarkable accuracy~\cite{Planck2018,ACT1,ACT2,ACT3,SPT}. The detailed pattern of acoustic peaks places stringent constraints on cosmological models, especially on the behavior of the metric potentials that drive the oscillations of the tightly coupled baryon--photon fluid prior to recombination.

In the standard $\Lambda$CDM cosmology, the persistence of these gravitational potentials is maintained by cold dark matter (CDM). Since CDM behaves as a collisionless, pressureless fluid with negligible sound speed, it clusters after horizon entry and prevents the rapid decay of the metric potentials during the radiation-dominated epoch. These nearly time-independent potential wells provide the driving term for the baryon--photon acoustic oscillator and generate the observed sequence of peaks in the CMB angular power spectra.

Despite the empirical success of this framework, the physical nature of dark matter remains unknown. Extensive laboratory searches have not produced direct evidence for a dark matter particle, and cosmological evidence for dark matter is inferred through its gravitational effects. This motivates the study of alternative gravitational theories in which the phenomena usually attributed to dark matter arise from additional gravitational degrees of freedom rather than from a new particle sector.

Scalar--Tensor--Vector Gravity (STVG), also known as Modified Gravity (MOG), is a covariant theory in which gravitation is mediated not only by the metric $g_{\mu\nu}$ but also by scalar fields and a massive vector field $\phi_\mu$~\cite{MoffatJCAP2006,BrownsteinMoffat2007,
MoffatToth2007,MoffatRahvar2013,MoffatRahvar2014,Moffat2013,Moffat2014,MoffatToth2021,GreenMoffat2019,GreenMoffatToth2018,DavariRahvar2020,Moffat2026}. In weak fields the theory yields a scale-dependent effective gravitational coupling $G_{\rm eff}$ and a repulsive vector contribution, and it has been successfully applied to galaxy rotation curves, cluster dynamics, and other astrophysical observables without invoking particle dark matter.

A central question is whether STVG--MOG can also account for the highly constraining CMB acoustic power spectra. The present paper argues that it can, provided two physical conditions are satisfied during the pre-recombination era. First, the massive vector field must admit nonrelativistic excitations that behave as a collisionless pressureless component with vanishing sound speed. Second, on the scales relevant for the acoustic peaks, the effective gravitational coupling must remain very close to Newton's constant, $G_{\rm eff}(k,a)\simeq G_N$. Under these conditions, the scalar perturbation dynamics become degenerate with those of $\Lambda$CDM through recombination.

The key observable is the third acoustic peak. In a universe containing only the baryon--photon fluid, radiation pressure causes the metric potentials to decay after horizon entry during radiation domination, suppressing the higher compressional peaks. The observed third peak requires a clustering, pressureless component that is already present before recombination and can sustain the gravitational potential wells on subhorizon scales. In the standard cosmological model, this role is played by CDM. In STVG--MOG, the same dynamical role can be supplied by nonrelativistic excitations of the vector field $\phi_\mu$, provided that their background density scales as $a^{-3}$ and their perturbations cluster as dust.

Once these conditions hold, the evolution of the metric potentials is the same as in General Relativity with CDM. The baryon--photon fluid then experiences the same acoustic driving, the same baryon loading, and the same diffusion damping as in $\Lambda$CDM. The positions and relative heights of the temperature and polarization peaks are consequently preserved. The important point is that the CMB acoustic spectra do not directly test the existence of a particle dark matter species; they test the existence of a gravitationally clustering, effectively pressureless component capable of sustaining the primordial potential wells through recombination.

An important conceptual issue concerns the interpretation of this effective dust component. In $\Lambda$CDM, the relevant component is an independent matter species whose density contributes to the stress-energy tensor on the right-hand side of the Einstein equations. In STVG--MOG, the corresponding component arises from excitations of a field belonging to the gravitational sector of the theory itself. The early-universe perturbation dynamics can be degenerate with those of CDM, while the component's physical origin remains fundamentally different. This distinction is important when one considers the interpretation of cosmological observables and the relation between modified gravity and particle dark matter.

A related discussion of the near-degeneracy between STVG--MOG and $\Lambda$CDM for isotropic linear cosmological observables was given in Ref.~\cite{Moffat2026}. The purpose of the present paper is narrower and more physical. We isolate the mechanism by which the STVG vector sector can supply the effective collisionless pressureless component required prior to recombination, and we identify the conditions under which the pre-recombination scalar sector becomes degenerate with $\Lambda$CDM. Our focus is on the physical origin of the degeneracy and the special role of the third acoustic peak as the decisive observational discriminator. The Boltzmann code CLASS is used to obtain a MOG fit to the acoustical power spectrum data.

The paper is organized as follows. In Section~2, we formulate the STVG--MOG perturbation framework relevant for the CMB and describe the nonrelativistic vector-sector excitations that act as an effective dust component. We show how the resulting metric potentials preserve the acoustic peak structure of the temperature and polarization spectra. In Section~3, we discuss why the vector-sector dust is not equivalent to particle dark matter, even though the linear perturbation dynamics in the early universe can be degenerate with those of $\Lambda$CDM. Section~4 contains our conclusions.

\section{CMB Acoustic Power Spectrum in STVG--MOG}
\label{sec:CMB_STVG}

The angular power spectra of the cosmic microwave background (CMB) temperature and polarization anisotropies provide one of the most stringent tests of the dynamics of cosmological perturbations. In the standard $\Lambda$CDM framework, the persistence of the gravitational potentials prior to recombination is
maintained by cold dark matter (CDM), which behaves as a collisionless, pressureless component with negligible sound speed. In the Scalar--Tensor--Vector--Gravity (STVG--MOG) framework, the same dynamical role is supplied by
nonrelativistic excitations of the vector field $\phi_\mu$.

In the early universe, the massive vector field admits nonrelativistic excitations whose background density evolves as~\cite{Moffat2014}:
\begin{equation}
\bar{\rho}_\phi \propto a^{-3}.
\end{equation}
The vector sector behaves as an effectively pressureless fluid:
\begin{equation}
p_\phi = 0, \qquad c_{s,\phi}^2 = 0, \qquad \sigma_{\phi}=0,
\end{equation}
where $p_\phi$, $c_{s,\phi}$ and $\sigma_{\phi}$ are the pressure, sound speed, and anisotropic stress, respectively. The perturbations obey the standard dust equations:
\begin{align}
\dot{\delta}_\phi &= -\theta_\phi + 3\dot{\Phi}, \\
\dot{\theta}_\phi + \Psi\theta_\phi &= k^2\Psi ,
\end{align}
where $\delta_{\phi} \equiv \delta \rho_{\phi}/\bar{\rho}_{\phi}$, and $\theta_\phi$ denote the density and velocity divergence perturbations, and $\Phi$ and $\Psi$ are the Newtonian-gauge metric potentials. Because anisotropic stress vanishes, $\Phi=\Psi$. The property of this component is that it is collisionless and pressureless, so that it clusters on subhorizon scales and sustains the gravitational potential wells during the epoch when the baryon--photon fluid is tightly coupled.

The weak-field STVG theory introduces a scale- and epoch-dependent effective gravitational coupling $G_{\rm eff}(k,a)$. During the pre-recombination era, the relevant cosmological scales satisfy~\cite{Moffat2014}:
\begin{equation}
G_{\rm eff}(k,a) \simeq G_N .
\end{equation}
The gravitational dynamics governing the growth of perturbations coincide with those of standard General Relativity. The Poisson constraint for the metric potential takes the form:
\begin{equation}
k^2 \Phi = -4\pi G_N a^2
\left(\rho_b \delta_b + \rho_\phi \delta_\phi + \rho_\gamma \delta_\gamma
+ \rho_\nu \delta_\nu \right).
\end{equation}

The background density of the vector sector satisfies:
\begin{equation}
\Omega_\phi(a_{\rm rec}) \simeq \Omega_{\rm CDM}(a_{\rm rec}).
\end{equation}
The evolution of the metric potentials is essentially identical to that in $\Lambda$CDM.

A critical requirement for reproducing the observed acoustic peak structure is that the gravitational potential remain sufficiently deep when the relevant
Fourier modes enter the horizon. In a universe containing only the baryon--photon fluid, radiation pressure would cause the potentials to decay after horizon entry during radiation domination. The presence of the pressureless vector component prevents this decay.

For modes satisfying: $k \sim aH $, where $H$ is the Hubble expansion rate, the vector dust clusters and sustains the gravitational wells through the Poisson equation. The resulting potentials remain nearly constant across the horizon-crossing epoch, ensuring the correct driving of the acoustic oscillator. This mechanism preserves the amplitude of the higher acoustic peaks, particularly the third peak, which is highly sensitive to the
persistence of the gravitational potential.

The baryon--photon perturbations satisfy the oscillator equation:
\begin{equation}
\ddot{\delta}_{\gamma b} + c_s^2 k^2 \delta_{\gamma b}
= -k^2 \Phi .
\end{equation}
The sound speed is given by
\begin{equation}
c_s^2 = \frac{1}{3(1+R)}, \qquad
R = \frac{3\rho_b}{4\rho_\gamma}.
\end{equation}

Since the metric potential $\Phi$ evolves in the same manner as in $\Lambda$CDM, the phase and amplitude of the acoustic oscillations remain unchanged. Consequently, the positions and relative heights of the acoustic
peaks in the angular power spectrum are reproduced.

The temperature anisotropy spectrum $C_\ell^{TT}$ and the polarization spectra $C_\ell^{EE}$ and $C_\ell^{TE}$ depend primarily on the evolution of the
metric potentials and the baryon--photon microphysics during recombination. Because STVG--MOG retains the standard recombination physics, Thomson scattering, photon diffusion, Silk damping, and baryon loading, the
predicted spectra satisfy:
\begin{equation}
C_\ell^{TT,EE,TE}({\rm STVG})
\simeq
C_\ell^{TT,EE,TE}(\Lambda{\rm CDM})
\end{equation}
provided that the vector sector supplies the effective dust component and
$G_{\rm eff} \approx G_N$ during the pre-recombination epoch.

\begin{figure}
    \centering
    \includegraphics[width=0.9\linewidth]{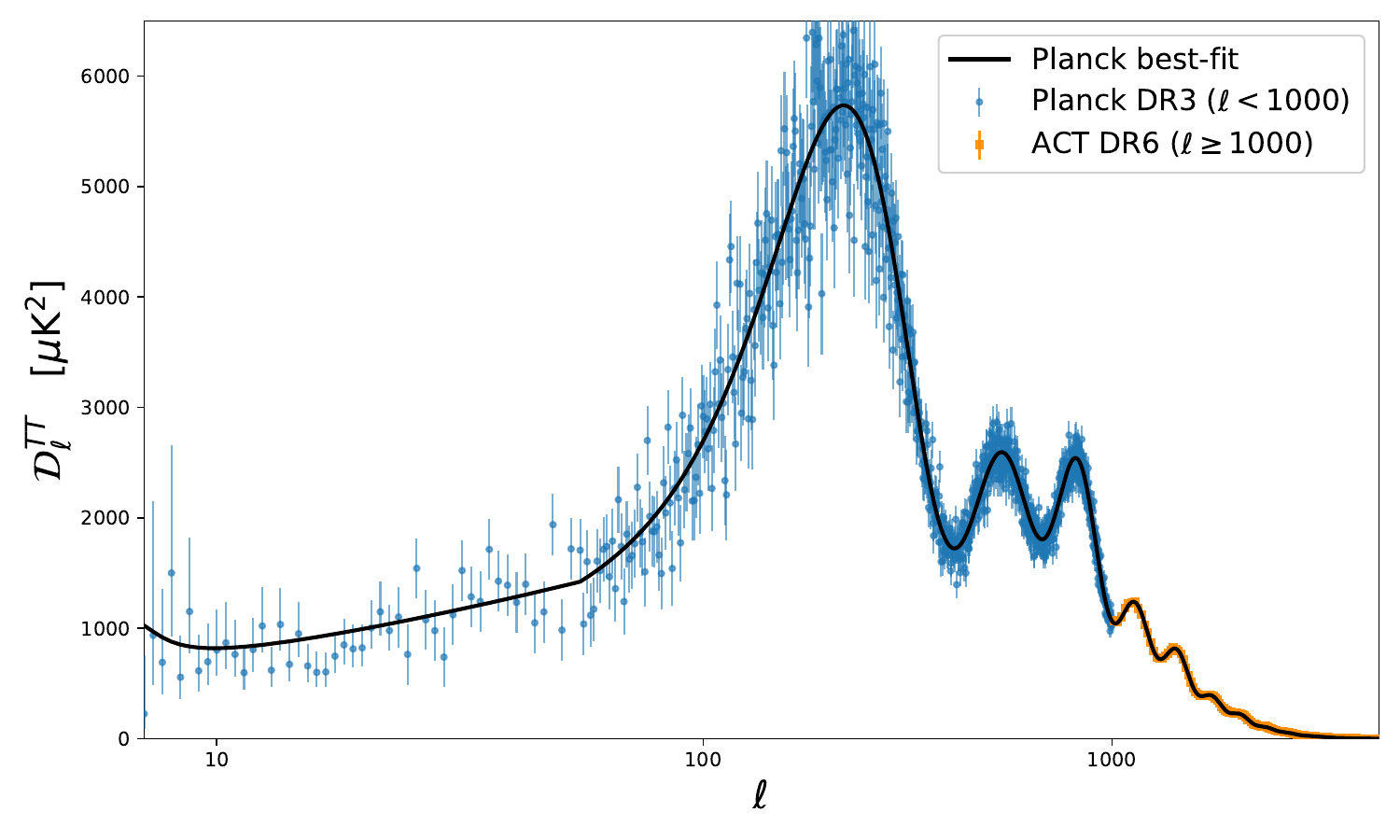}
    \caption{CMB temperature anisotropy acoustic power spectrum in STVG--MOG compared with the observed TT data.  The plotted quantity is
\(D_\ell^{TT}\equiv \ell(\ell+1)C_\ell^{TT}/(2\pi)\).  }
\label{fig:mog-tt-spectrum}
\label{fig:placeholder}
\end{figure}

\section{Difference of STVG-MOG from Particle Dark Matter}

In this framework, the dynamics normally attributed to particle dark matter are instead generated by a purely gravitational degree of freedom. The vector-field excitations act as a collisionless pressureless component that stabilizes the metric potentials and reproduces the observed acoustic peak structure of the CMB without introducing a new particle sector.

Because the nonrelativistic excitations of the STVG vector field behave as a collisionless pressureless component during the pre-recombination epoch, it is natural to ask whether this component is simply equivalent to cold dark matter (CDM). Although the perturbative dynamics can be identical in the early universe, the physical origin of the component and its role in the gravitational theory are fundamentally different.

In the $\Lambda$CDM framework, CDM represents a new particle species belonging to a matter sector beyond the Standard Model. Its energy density is an independent component of the cosmic stress-energy tensor
and its dynamics are governed by particle physics properties such as mass, interactions, and phase-space distribution. The CDM density appears in the Einstein field equations through the matter stress tensor:
\begin{equation}
G_{\mu\nu} = 8\pi G_N
\left(T_{\mu\nu}^{\rm baryon}
+ T_{\mu\nu}^{\rm radiation}
+ T_{\mu\nu}^{\rm CDM}
+ \cdots \right).
\end{equation}

By contrast, in STVG--MOG, the vector field $\phi_\mu$ is not a new matter species, but a gravitational degree of freedom that is part of the fundamental gravitational sector of the theory. The action contains the vector field together with the metric and scalar fields:
\begin{equation}
S = S_{\rm GR}
+ S_\phi
+ S_{\rm scalar}
+ S_{\rm matter},
\end{equation}
where $S_\phi$ describes the dynamics of the massive vector field that modifies the gravitational interaction. The nonrelativistic excitations of $\phi_\mu$ arise from this gravitational sector rather than from a new particle sector.

The dust-like behavior of the vector excitations is a
description valid in the linear perturbation regime of cosmology in the early universe. The repulsive vector field associated with $\phi_\mu$ density satisfies $\rho_\phi >> \rho_b$. These excitations cluster gravitationally and sustain the metric potentials prior to recombination, thereby reproducing the same acoustic peak structure that CDM produces in the standard cosmology. However, their origin is entirely different; they are fluctuations in a gravitational field rather than particles of dark matter. This distinction becomes particularly important at late times and on astrophysical scales. In STVG--MOG, the gravitational interaction at late time is modified through the scale-dependent effective coupling $G_{\rm eff}(k,a)$, and in the baryon dominated late time universe the vector field density is smaller than the baryon density, $\rho_\phi << \rho_b$. Galaxy rotation curves, cluster dynamics, and large-scale structure can be described without invoking a massive halo of dark matter. In the early universe, the vector field density $\rho_\phi >> \rho_b$ in contrast to the late time baryon dominated universe. In the standard cosmological model, by contrast, these phenomena require a dominant CDM component at every epoch of the universe.

While the early-universe perturbation dynamics can mimic those of $\Lambda$CDM, the physical interpretation is fundamentally different. In STVG--MOG, the component responsible for sustaining the primordial gravitational potential is a gravitational field excitation rather than
a particle dark matter fluid.

\section{Conclusions}

The acoustic structure of the cosmic microwave background provides one of the most stringent tests of cosmological models. In the standard $\Lambda$CDM framework, the persistence of the gravitational potentials that drive baryon--photon oscillations prior to recombination is
maintained by cold dark matter, which behaves as a collisionless, pressureless component with negligible sound speed. The detailed heights and positions of the acoustic peaks encode direct information about the dynamical components that contribute to the gravitational potential wells in the early universe.

We have shown that Scalar--Tensor--Vector Gravity (STVG--MOG) provides an alternative mechanism that reproduces the observed CMB temperature and polarization power spectra without introducing particle dark matter. The key element is the existence of nonrelativistic excitations of the STVG vector field $\phi_\mu$ that behave as a collisionless pressureless component with vanishing sound speed. These excitations redshift as $\rho_\phi \propto a^{-3}$ and cluster gravitationally in the early universe in the same manner as a dust component.

On the cosmological scales relevant for the generation of the acoustic peaks, the effective gravitational coupling satisfies $G_{\rm eff}(k,a)\simeq G_N$. As a consequence, the evolution of the metric potentials governing the baryon--photon acoustic oscillations is identical to that in the standard cosmological model. Because the
vector-sector dust is present prior to the horizon entry of the relevant Fourier modes, the gravitational potentials remain sufficiently deep during the radiation-dominated epoch to preserve the observed height
of the third acoustic peak. Since recombination physics, Thomson scattering, baryon loading, and photon diffusion remain unchanged, the resulting temperature and polarization spectra $C_\ell^{TT}$, $C_\ell^{TE}$, and $C_\ell^{EE}$ coincide with those obtained in $\Lambda$CDM.

An important conceptual point is that the component responsible for sustaining the primordial gravitational potentials is not a new matter species. The nonrelativistic vector excitations arise from the gravitational sector of STVG itself rather than from a particle dark matter sector. Although the perturbation dynamics in the early universe can be degenerate with those of $\Lambda$CDM, the physical origin of the dust component is fundamentally different. In STVG--MOG, the phenomena normally attributed to dark matter emerge from additional gravitational degrees of freedom rather than from an unknown particle species.

The acoustic peak structure of the CMB does not uniquely demonstrate the existence of particle dark matter; instead, it probes the behavior of gravitational potentials in the early universe. The results presented here show that a purely gravitational component
arising from the STVG vector sector can provide the required dynamical support for these potentials while reproducing the observed acoustic power spectrum and polarization data.

Future work will involve implementing the full STVG perturbation framework in Boltzmann codes such as \texttt{CLASS} or \texttt{CAMB} to compute high-precision spectra and to confront the theory with the complete set of cosmological observations, including large-scale structure, lensing, and late-time cosmic expansion. Such analyses will allow a detailed assessment of the viability of STVG--MOG as an alternative gravitational framework capable of explaining cosmological data without particle dark matter.

\section*{Acknowledgments}

The author gratefully acknowledges Caio Nascimento for performing the Boltzmann code CLASS calculations that were used to obtain the MOG fit to the CMB acoustic power spectrum data. Research at the Perimeter Institute for Theoretical Physics is supported by the Government of Canada through Industry Canada and by the Province of Ontario through the Ministry of Research and Innovation (MRI).


\begin{thebibliography}{99}

\bibitem{Planck2018}
N.~Aghanim \textit{et al.}, (Planck Collaboration),
``Planck 2018 results. VI. Cosmological parameters,'' Astron.\ Astrophys.\ \textbf{641}, A6 , 67 (2020), arXiv:astro-ph.CO/1807.06209.

\bibitem{ACT1} Sigurd Naess \textit{et al}., ``ACT Collaboration, The Atacama Cosmology Telescope: {DR6} maps, ''JCAP, \textbf{11}, 061 (2025), arXiv:astro-ph.GA/2503.14451.
  
\bibitem{ACT2} T. Louis \textit{et al}., 
``ACT Collaboration, The Atacama Cosmology Telescope: {DR6} power spectra, likelihoods and {$\Lambda$CDM} parameters, ''JCAP, \textbf{11}, 062 (2025), arXiv: astro-ph.CO/2503.14452.
  
\bibitem{ACT3}  E. Calabrese, \textit{et al}., ``ACT Collaboration, The Atacama Cosmology Telescope: {DR6} constraints on extended cosmological models, ''JCAP,
\textbf{11}, 063 (2025), arXiv:astro-ph.CO/2503.14454.

\bibitem{SPT} E. Camphuis et al., 
``SPT-3G D1: CMB temperature and polarization power spectra and cosmology from 2019 and 2020 observations of the SPT-3G Main field", Phys. Rev. D 113, 083504 (2026).

\bibitem{MoffatJCAP2006}
J.~W.~Moffat, ``Scalar--Tensor--Vector Gravity Theory,''
JCAP \textbf{03}, 004 (2006), arXiv:gr-qc/0506021.

\bibitem{BrownsteinMoffat2007} J. R. Brownstein and J. W. Moffat, ``The Bullet Cluster 1E0657-558 evidence shows modified gravity in the absence of dark matter", MNRAS 382, 29 (2007).

\bibitem{MoffatToth2007} J.~W.~Moffat and V.~T.~Toth,
``Fundamental parameter-free solutions in Modified Gravity,'' Class. Quantum Grav. \textbf{26}, 085002 (2009), arXiv:0712.1796v5.

\bibitem{MoffatRahvar2013} J. W. Moffat and S. Rahvar, MNRAS 436, 1439 (2013).

\bibitem{MoffatRahvar2014} J. W. Moffat and S. Rahvar, MNRAS 1, 3724 (2014).

\bibitem{Moffat2013}
J.~W.~Moffat and V. T. Toth,
``Cosmological Observations in a Modified Theory of Gravity,''Galaxies \textbf{1}(1), 65 (2013),
arXiv:1104.2957.

\bibitem{Moffat2014}
J.~W.~Moffat,
``Structure Growth and the CMB in Modified Gravity (MOG),'' arXiv:1409.0853.

\bibitem{MoffatToth2021}
J.~W.~Moffat and V.~T.~Toth,
``Scalar--Tensor--Vector modified gravity in light of the Planck 2018 data,'' Universe, 7, 358 (2021), arXiv:2104.12806.

\bibitem{GreenMoffat2019}
M.~A.~Green and J.~W.~Moffat,
``Modified Gravity (MOG) fits to observed radial acceleration of SPARC galaxies,''
Phys.\ Dark Univ.\ \textbf{25}, 100323 (2019),
arXiv:1905.09476.

\bibitem{GreenMoffatToth2018}
M.~A.~Green, J.~W.~Moffat, and V.~T.~Toth,
``Modified Gravity (MOG), the speed of gravitational radiation and the event GW170817/GRB170817A,''
Phys.\ Lett.\ B \textbf{780}, 300--302 (2018),
arXiv:1710.11177.

\bibitem{DavariRahvar2020}
Z.~Davari and S.~Rahvar,
``Testing Modified Gravity (MOG) theory and dark matter model in Milky Way using the local observables,''
Mon.\ Not.\ R.\ Astron.\ Soc.\ \textbf{496}, 3502--3511 (2020).

\bibitem{Moffat2026}
J.~W.~Moffat,
``Isotropic equivalence of STVG--MOG and $\Lambda$CDM and its breakdown in large--scale anisotropic cosmological observables,'' arXiv:2601.22207 (v2).



\end{thebibliography}
\end{document}